\documentstyle[12pt,twoside,fleqn,espcrc1,epsfig]{article}

\newcommand{\AmS}{{\protect\the\textfont2
  A\kern-.1667em\lower.5ex\hbox{M}\kern-.125emS}}

\title{ Color Dynamics In Phase Space: The Balescu-Lenard-Vlasov Approach }

\author{A. Bonasera
\footnote{Bonasera@lns.infn.it}
\address{
Laboratorio Nazionale del Sud, Istituto Nazionale di Fisica Nucleare, 
Via S. Sofia 44, 95123 Catania, Italy
.
 }}
\begin{document}
\maketitle
\begin{abstract}
We propose to use the Balescu-Lenard-Vlasov (BLV)
 equation to describe relativistic heavy ion collisions. 
We use an inter-quark Richardson's potential
consistent with the indications of Lattice QCD calculations. The color
degrees of freedom are explicitly taken into account.  
 We explicitly demonstrate that the Vlasov approach alone is
insufficient in the hadronization region.  In order to overcome
this problem we prepare the initial condition for many events using molecular
 dynamics with frictional cooling and a Thomas-Fermi approximation to the
Fermi motion. 
  These events are averaged and propagated in time using the Vlasov
 approach.  The fluctuations are used to evaluate the collision term and in
turn the number of q-$\bar q$ created.
\end{abstract}



\section{INTRODUCTION}
One of the open problems in theoretical nuclear and particle physics is
how to obtain the well known nuclear properties starting from the
quark degrees of freedom \cite{pov}. This also includes the possibility of
understanding the basic free nucleon-nucleon interaction from quark and
gluon dynamics.  Some kind of solution to this problem is becoming more
and more needed with the new experiments done or planned using  
 ultra-relativistic heavy ions (rhic) at CERN and at RHIC.  
The search for a quark-gluon plasma (QGP) in such collisions is in fact 
one of the new and most exciting
directions in physics at the border between nuclear and particle physics
\cite{wong}. Quantum ChromoDynamics
(QCD) because of its difficulties (numerical and conceptual),
has been applied so far to some
limited cases such as quark matter at zero baryon ($\rho_ B$)
density and high temperature
(T) \cite{pov,wong}. Furthermore in rhic
dynamics plays surely an important role and accordingly the theory should 
 be dynamical.

 Recently \cite{bon99}, 
we have proposed a dynamical approach based on the
Vlasov equation \cite{repo,land1} to reproduce hadron masses and the properties
of nuclear matter at finite $\rho_B$. 
Some works in the same spirit are  discussed in \cite{mosel,horst}.
Our approach
needs as inputs the interaction potential among quarks, which was borrowed
from phenomenology i.e. the Richardson's potential\cite{rich}, and the quark 
masses which were fitted to reproduce 
known meson masses such as the $\pi$, the $\phi$, the 
$\eta_c$ etc.  When the particles are embedded in a dense medium such as
in nuclear matter (NM) the potential becomes screened in a similar fashion
as ions and electrons in condensed matter do, i.e. 
Debye screening (DS)\cite{wong,land1}. 

Here we  refine that approach in two important
aspects which are the treatment of the color degrees of freedom and the
inclusion of the collision term.  In the
previous works \cite{bon99} color degrees of freedom were implicitly
taken into account through the use of a Debye radius that effectively
screens the $qq$ interaction potential.  In the present paper we 
 give to the quarks explicitly a color (using the Gell-Mann matrices)
and follow their dynamics in phase space solving the Vlasov equation (VE).
 Thus screening will be dynamically obtained.  
In general the self screening
 obtained in Vlasov dynamics is inadequate, which is the reason why it was not
adopted in the earlier attempts \cite{bon99}.  In fact we will
 show explicitly that the Vlasov approach alone gives a good 
description of the system at large densities only,i.e. in the
 QGP region.  In order to overcome this
problem we adopted the following strategy. 
We first prepare the initial conditions using molecular dynamics (MD)
with frictional cooling for many events.  The events are averaged
and care is taken of antisymmetrization. These are the initial conditions
for the Vlasov evolution.  Since the VE fulfills the Liouville theorem,
the initial phase space density remains constant in time. Thus the initial
antisymmetrization and eventual clustering obtained in the cooling process
are maintained. When simulating a nucleus-nucleus collision, inter quark
 correlations and particle production become very important.  We have
included this feature using the most general form of the collision term
which is the Balescu-Lenard type.  This is very crucial since we are dealing
with particles which are interacting through long range forces. 

\section{The Balescu Lenard Vlasov equation}
We outline our approach on purely classical grounds, however
the same results can be obtained within the Wigner transform formalism
\cite{repo,land1} of the quantum BBGKY-hierarchy in 
the limit $\hbar \rightarrow 0$.

The exact (classical) one-body distribution function $f_1(r,p,t)$ 
 satisfies
the equation (BBGKY hierarchy)\cite{land1}:
   \begin{equation}
\partial_{t}f_1+\frac{\overrightarrow{p}}{E}\cdot \nabla _{r}f_1
=\int{d(2)\nabla_r V({\bf r},{\bf r_2})\nabla_p f_2({\bf r,r_2,p,p_2},t)}
\end{equation}
$E=\sqrt{p^2+m_i^2}$ is the energy and 
$m_i=10 MeV$ is the (u,d) quark mass. 
Here we assume the potential to be dependent on the relative coordinates only.
A generalization to include a momentum dependent part is straightforward.
 $f_2$ is the two-body distribution function, which in the classical limit
reads:
\begin{equation}
f_2(r,r_2,p,p_2,t)=\Sigma_{\alpha \ne \beta}^{Q}\delta ( \bf r- \bf r_{\alpha})
\delta (\bf p- \bf p_{\alpha}) \times \delta ( \bf r_2- \bf r_{\beta})
\delta ( \bf p_2- \bf p_{\beta})
\label{lv2}
\end{equation}
where $Q=q+\bar q$ is the total number of quarks and anti quarks
.  
Inserting this equation into Eq.(1) gives:
\begin{equation}
\partial_{t}f_1+\frac{\overrightarrow{p}}{E}\cdot \nabla _{r}f_1
-\nabla_{r}U\cdot \nabla _{p}f_1=0
\label{lv3}
\end{equation}

Where $U=\Sigma_j V(\bf r,\bf r_j)$ is the exact potential.  Let us now
define $f_1$ and $U$ as sums of an ensemble averaged quantity plus the
deviation from this average:
\begin{equation}
f_1=\bar f_1+\delta f_1;  
U=\bar U+\delta U
\label{lv4}
\end{equation}
Substituting these equations in Eq.(3) and ensemble averaging gives:
   \begin{equation}
\partial_{t}\bar f_1+\frac{\overrightarrow{p}}{E}\cdot 
\nabla _{r}\bar f_1
-\nabla
_{r}\bar U\cdot \nabla _{p}\bar f_1= 
<\nabla_r\delta U\nabla_p\delta f_1>
\label{lv}
\end{equation}

where one recognizes in the lhs the Vlasov term and in the rhs the
Balescu-Lennard collision term \cite{land1,belk,crist}.
The mean-field is given by:
    \begin{equation}
\bar U(\bf r)=\frac{1}{N_{ev}}\Sigma_{ev}\Sigma_j V(\bf r,\bf r_j)
\end{equation}
In agreement to LQCD calculations\cite{pov,rich} the interacting potential
$V(r)$ for quarks is ($\hbar=1$):
 
 \begin{eqnarray}
 V(r_{i,j})=3\Sigma_{a=1}^{8}\frac{\lambda_i^a}{2}\frac{\lambda_j^a}{2}[
\frac{8\pi}{33-2n_f}\Lambda(\Lambda r_{ij}-\frac{f(\Lambda r_{ij})}
{\Lambda r_{ij}})
+\frac{8\pi}{9}
\bar{\alpha}\frac{<{\bf \sigma}_q{\bf \sigma}_{\bar q}>}
{m_q m_{\bar q}}\delta({\bf r_{ij}}) ]
 \end{eqnarray}
 
 and\cite{rich}
 \begin{eqnarray}
f(t)=1-4 \int{\frac{dq}{q}\frac{e^{-qt}}{[ln(q^2-1)]^2+\pi^2}} 
 \end{eqnarray}

We fix the number of flavors 
$n_f=2$ and
the parameter $\Lambda=0.25 GeV$ 
.
In Eq.(7) we have added to the Richardson's potential 
the chromomagnetic term (ct), very important
 to reproduce  the masses of
the hadrons in vacuum.  
Since in this work we will be dealing with finite nuclei, the ct can
 be neglected, we only notice that with the parameters choice discussed here,
the hadron masses can be reproduced by suitably tuning the ct term
\cite{bon99}. 

  The $\lambda^a$ are the Gell-Mann matrices.
  From lattice calculations we expect that there is no color transport for
distances of the order of $0.2-0.3 fm$, which are distances much shorter
than the ones we will be dealing with in this paper. Thus we will use
 the $\lambda_{3,8}$ only commuting diagonal Gell-Mann matrices (Abelian
 approximation)\cite{horst}.
 
Numerically the BLV equation(5) is solved by writing the one body distribution
function as:

\begin{equation}
\label{efer}
\bar f_1(r,p,t) = \frac{1}{n_{tp}} \sum_i^N 
\sigma_i(t) \delta(r-r_i(t)) \delta(p-p_i(t)) 
\end{equation}

 $N=Qn_{tp}$ is the number of such terms.
  Actually, 
N is much larger than the total quark
number Q, so that we can say that each quark 
is represented by ${n_{tp}}=N_{ev}$ terms called test particles(tp).
 Inserting Eq.(9) in the BLV  equation 
 gives
the Hamilton equations of motion (eom) for the tp~\cite{repo}. The 
weight factor fulfills the equation $
\partial_{t}\sigma_i(t)=
\nabla_{r_i}\delta U\nabla_{k_i}\delta f_1$ \cite{prl93}. 
 Initially the weight factor is equal
to 1 for the tp.  When for a particular tp the weight factor becomes larger
than 2 a new tp (quark) is created in the same phase space region.  In the
same way if the weight becomes less than zero for a given tp another tp
with negative sign ( an anti-quark) is created.  Notice that on average
 over time and ensembles the number of newly created quarks is equal to the
number of anti quarks.  Particular care is taken in order to have specific color 
conservation as well. We would like to stress now that in our picture the mean
 field and the collision term are calculated in a consistent way starting from
a two body potential.  This is indeed a quite general method which can be
applied to other physical systems.
%
%

Initially, 
 we distribute randomly  the tp in a sphere of radius $R=r_{0B}A^{1/3}$ 
(the radius of the nucleus) in coordinate
space and $p_f$ in momentum space.  $r_{0B}=(\frac{3}{4\pi\rho_B})^{1/3}$,
$A=Q/3$ and $\rho_B$ is the baryon density.
$p_f$ is the Fermi momentum
estimated in a simple Fermi gas model by imposing that a cell in
phase space of size $h=2 \pi$ can accommodate at most two identical quarks
 of different spins, flavors and colors. A
simple estimate gives the following relation between the quark density
 ,
$n_{q}$, and the Fermi momentum:
\begin{eqnarray}
n_{q}=\frac{g_q}{6\pi^2}p_f^3 
 \end{eqnarray}
  The degeneracy number 
$g_q=n_f\times n_c\times n_s$, where $n_c$ is the number of colors and
$n_s$ is the number of spins\cite{wong}.  For quarks and
 anti quarks 3 different colors are used red, 
green  and blue (r,g,b) \cite{pov}.  

\section{Results}
\begin{figure}[htp]
\begin{minipage}[t]{150mm}
\epsfysize=8.8cm
\centerline{\epsfbox{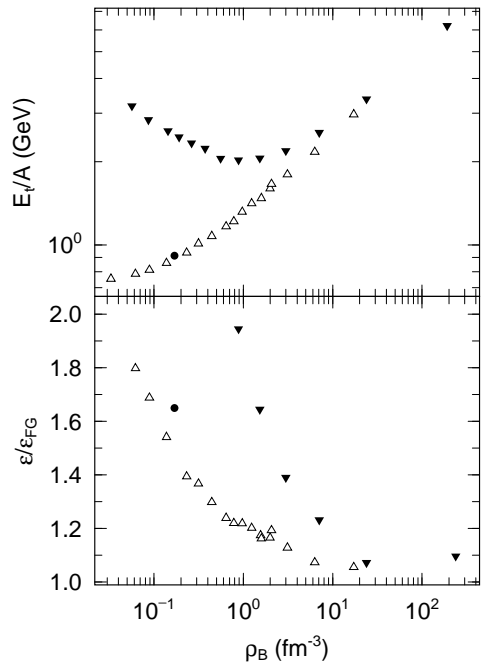}}
\caption{Energy per nucleon (top) and energy density
of the quarks 
(bottom) vs. baryon density.  
} 
\end{minipage}
\end{figure}
In figure(1), we plot the total energy per nucleon (top) and energy
 density (in units of the Fermi gas energy density\cite{wong}) vs.
baryon density. The full triangles give the results
obtained by randomly distributing the tp as described above.  We notice
that  a minimum at about $\rho_c=2.08 fm^{-3}$ is found with $E_t/A=2GeV/A$.  
Such a minimum is at much higher density and energy than expected for
the ground state (gs) of a nucleus 
($\rho_0=0.16 fm^{-3}$ and $E_t/A=0.938-0.016 GeV/A$).
An important property of the system that we have  described
above is the following.  If we rotate the quarks in color space,
regardless of their position in r-space, the
total energy will remain the same. 
This is indeed a "pure" Vlasov solution and demonstrates explicitly the
in-capability of the Vlasov approach to give clustering of quarks into
 nucleons.  However this result is already instructive since it gives us an
 hint on where the quark and nuclear matter are located, i.e. above and below
$\rho_c$ respectively. This result is qualitatively in agreement
with the Hartree-Fock(HF) calculations of refs.\cite{horo,roe}
(compare to fig.1 in \cite{horo}).
For a discussion on why the $E_t/A$ increases at low densities in the HF/VE
 approaches we also refer to \cite{horo}. 

  Of course distributing randomly
the quarks in a sphere in r and p-space is not the most economical way
 to find the real gs of the system.  In MD one 
searches for a minimum energy by introducing a friction term.  
  The friction acts in such a way to lead the
 particles to a configuration for which the potential energy is a minimum.
 We cannot use this technique for our system since we are dealing with
 fermions and the friction term will destroy the initial 
antisymmetrization.  In order to overcome this difficulty we adopt the
following strategy.  First we prepare $N_{ev}$ events as above, and
we evolve them numerically solving the  
eom but with  friction included. 
Because of the large number of particles interacting with attractive 
and repulsive forces, the quarks will slowly evolve to new
 positions where the potential is lower while keeping the initial
root mean square radius approximately constant.  When the averaged potential 
(over events) reaches a given value $V_{min}$, 
we look for the two closest particles $\it (j,k)$ to
a quark $\it i$ in the same event.  
For these three quarks we know what the local
density and the number of colors are.  For instance if in a certain region
we find two red and one blue quark,  we use $n_c=2$ in Eq.(10) and
calculate 
the local density from the knowledge of the distances of the 3 quarks.  
In this way the Fermi momentum is defined locally
similar to the procedure used in nuclear or atomic physics (Thomas-Fermi
approximation)\cite{schuck}. We repeat this for all quarks $\it i$
 in all events and calculate
 the total energy for this state
.  We let the system evolve again with friction included 
to a lower potential energy $V_{min}=V_{min}-\delta V$, where $\delta V$ is a 
constant. We calculate the local density and local color numbers again
and apply the Thomas-Fermi approximation to obtain the new total energy.  
We stop the procedure when
the total energy is a minimum. The initial conditions so obtained are
 then propagated in time using the VE in order to maintain the initial
 antisymmetrization. 

The open triangles in fig.(1) are the
result of the minimization.  Notice especially at low densities the
large decrease of the total energy of the system as compared to the
Vlasov result.  Now the calculated 
total energy
at the nuclear gs is very close to the experimental value indicated by the
full circle. However, we find slightly lower energies at lower densities, 
 i.e. the gs of the nucleus is shifted in our calculations at about
1/10 of the experimental one.  This should not be surprising in view of 
the simple potential that we have used. Also we have not tried to
look for a best set of fitting parameters in these exploratory studies.  
At low $\rho_B$,
 the global invariance for rotations in color space
is lost, i.e. if we exchange the colors of two ${\it distant}$ quarks, 
the total
energy of the system will change.  
  At high densities (larger
than $2 fm^{-3}$) 
the Vlasov and MD solutions are the same.
  This can be also seen in the bottom part of the figure where energy
densities are given. We would like to stress again the qualitative agreement 
to ref.\cite{horo} where a stochastic method had been used to calculate the
g.s. energy of the system.  This is quite evident if one compares our fig.1
 to fig.1 of \cite{horo}.

The energy density displayed in fig.1 (bottom) is a smooth
function apart some fluctuations around $2 fm^{-3}$ density.  From
this result we can exclude a first order phase transition but a second
order phase transition might be possible, see the discussion in \cite{bon00}.

\begin{figure}[htp]
\begin{minipage}[t]{150mm}
\epsfysize=14.8cm
\centerline{\epsfbox{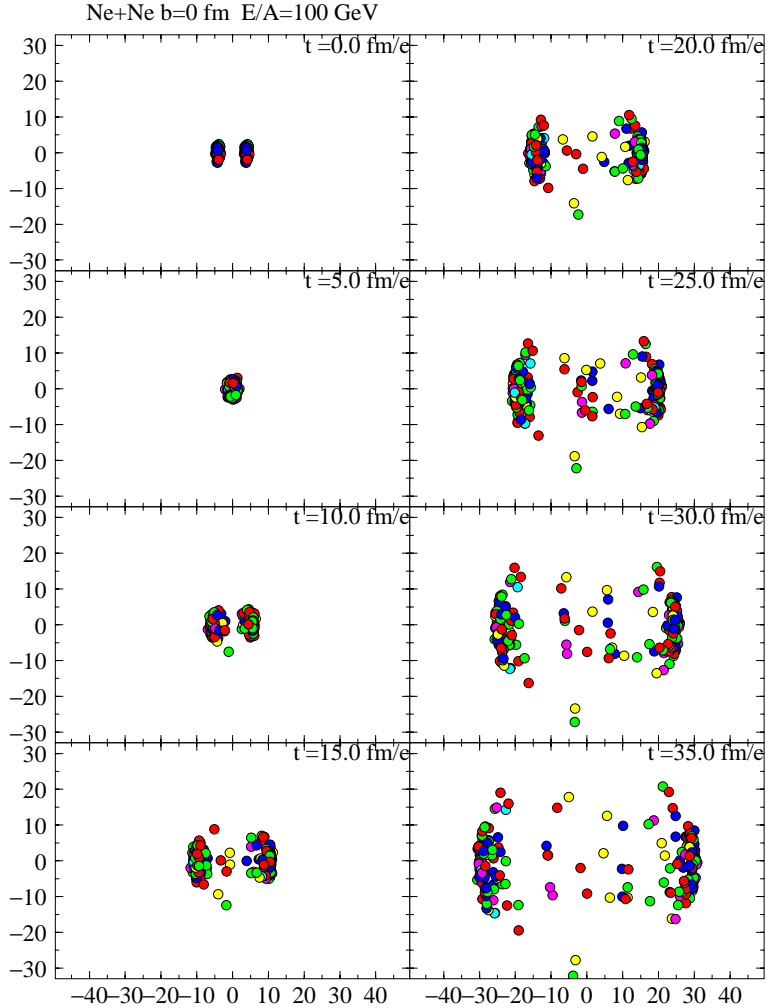}}
\caption{ Ne+Ne collision at 100 GeV/A and b=0fm. The quarks are given by
the (r,g,b) while the created anti quarks by the yellow, magenta and cyan
colors.
 }
\end{minipage}
\end{figure}
Once we have devised a method to find a (approximate) 
nucleus ground state starting from quark degrees of freedom we can perform
nucleus-nucleus collision simulations. In figure (2) we plot snapshots of
a simulation for Ne+Ne collision at 100 GeV/A.  Initially the quarks are
distributed randomly in a sphere of radius R and the minimization procedure
is applied.  Such an initial condition is Lorentz boosted and the dynamics
is followed in time.  In this simulation $N_{ev}=9$ for numerical reasons and
one event only is displayed in the figure (2).  Because of the large 
fluctuations from event to event, the collision term is very important and
 new quarks and anti quarks are created.  These are displayed in the figure
with different colors.  The newly created quarks are strongly sensitive to
the parameters entering the two body force and especially to the linear
term.  This might have some connection with the Schwinger particle 
production mechanism \cite{wong} but more studies are needed to find
 a precise link (if any).

\section{Summary}
In conclusion in this work we have discussed  microscopic Vlasov/MD 
approaches
to finite nuclei starting from quark degrees of freedom with colors.
In order to obtain the correct initial conditions we have introduced a method
based on MD with frictional cooling plus a Thomas-Fermi approximation for
the Fermi motion.  
We have shown that the method is able to describe at least qualitatively
 the well known features of nuclei near the ground state.  At high
densities a second order phase transition from nuclear to quark matter
is predicted.
  Such a transition is due to the restoration of global
color invariance at high densities and we have defined an order parameter
accordingly\cite{bon00}. 
We have  used the initial conditions for the nucleus ground state to 
simulate heavy ion collisions at ultra-relativistic energies with the
Balescu Lenard 
collision term.  The mean field and the collision term are calculated 
starting from the same elementary interactions. Collisions are 
connected to deviations from the average mean field.  This gives
 also large fluctuations in the distribution function which results in 
new particles creation. Our approach can be
very useful for the understanding of the quark gluon plasma formation and
its signatures.

I thank Toshiki Maruyama for useful discussions, suggestions and the use
of drawing subroutines.  



%
%
%
%

%
\hspace{\fill}
%

\end{document}